\documentclass{article}[11pt]
\usepackage{amsfonts}
\usepackage{booktabs}
\usepackage{graphicx}
\usepackage{CJK,amsmath,indentfirst}
\usepackage{amssymb}
\usepackage{amsmath}
\usepackage{amsthm}
\usepackage{epsfig}
\usepackage{times}
\usepackage{subfigure}
\usepackage[numbers,sort&compress]{natbib}
\begin{document}

\title{\bf Nonlocal Symmetries and Interaction Solutions of the (2+1)-dimensional Higher Order Broer-Kaup System }

\author{\footnotesize  Xiangpeng Xin  \thanks{Corresponding author ~~~~ Email:xinxiangpeng2012@gmail.com} \\
\footnotesize  School of Mathematical Sciences, Liaocheng University, Liaocheng 252059, People's Republic of China}
\date{}
\maketitle
\parindent=0pt
\textbf{Abstract:}  The (2+1)-dimensional higher-order Broer-Kaup (HBK) system is studied by nonlocal symmetry method and consistent tanh expansion (CTE) method in this paper. Some exact interaction solutions among different nonlinear excitations such as solitons, rational waves, periodic waves and corresponding images are explicitly given. 

\vskip.4in
\renewcommand{\thesection}{\arabic{section}}
\parindent=20pt

\section{Introduction}

Finding explicit solutions of nonlinear partial differential equation(NPDE)is one of the most important problems in mathematical physics.
With the development of nonlinear science, many methods have been established by mathematicians and physicists to study the integrability of NPDEs. Such as inverse scattering transformation\cite{Gardner1,Ablowitz1}, Painlev\'{e} analysis\cite{Weiss1,Conte1}, classical and non-classical Lie group\cite{Olver1,Bluman1}, nonlocal symmetry\cite{Lou1,Galas1,Hu1}, variable separation approach\cite{Lou2,Lou3}, and various function expansion methods\cite{Fan1,Fan2} etc. Recently, many scholars have studied the problem of interaction solutions to NPDEs\cite{Wang1,Hxr1}, however, it is very difficult to find interaction solutions among different types of nonlinear excitations except for soliton-soliton interactions.

It is known that Painlev\'{e} analysis is an important method to investigate the integrable property of a given NLEE, and the truncated Painlev\'{e} expansion method is a straight way to provide auto-B\"{a}cklund transformation and analytic solution, furthermore, it can also be used to obtain nonlocal symmetries. Recently, by developing the truncated Painlev\'{e} expansion, Lou\cite{Lou4,Lou5,Lou6,Lou7} defined a new integrability in the sense of possessing a consistent tanh expansion. This method is greatly valid for constructing various interaction solutions between different types of excitations. For example, solitons, cnoidal waves, Painlev\'{e} waves, Airy waves, Bessel waves etc. It has been revealed that many more integrable systems are CTE solvable and posses quite similar interaction solutions which can be described by the same determining equation with different constant constraints.

This paper is arranged as follows: In Sec.2, from the truncated Painlev\'{e} expansion, the residual symmetry of the HBK system is obtained, and the nonlocal symmetry group is found by the localization process. In Sec.3, Using the CTE method, some interaction solutions between different types of excitations and corresponding images are given. Finally, some conclusions and discussions are given in Sec.4.

\section{Nonlocal symmetries of (2+1)-dimensional HBK System}

In this section, we concentrate on investigating the nonlocal symmetries of HBK system. The 2+1 dimensional HBK system\cite{Lin1,Li1},
\begin{equation}\label{bk-1}
\begin{array}{l}
 u_{yt}  + 4(u_{xx}  + u^3  - 3uu_x  + 3uw)_{xy}  + 12(uv)_{xx}  = 0, \\
 v_t  + 4(v_{xx}  + 3vu^2  + 3uv_x  + 3vw)_x  = 0, \\
 w_y  - v_x  = 0, \\
 \end{array}
\end{equation}
where $u=u(x,y,t),v=v(x,y,t),w=w(x,y,t)$.

This system was first obtained from the inner parameter dependent symmetry constraints of the KP equation. When we take $y=x$, the system (\ref{bk-1}) is reduced to the usual (1+1)-dimensional HBK system.

For the HBK system(\ref{bk-1}), we take a truncated Painlev\'{e} expansion,
\begin{equation}\label{bk-2}
u = \frac{{u_0 }}{\phi } + u_1 ,v = \frac{{v_0 }}{{\phi ^2 }} + \frac{{v_1 }}{\phi } + v_2 ,w = \frac{{w_0 }}{{\phi ^2 }} + \frac{{w_1 }}{\phi } + w_2 ,
\end{equation}
with $u_0,u_1,v_0,v_1,v_2, w_0,w_1,w_2,\phi$ being the functions of $x, y$, and $t$.

Substituting Eqs.(\ref{bk-2}) into system(\ref{bk-1}) and vanishing all the coefficients of different powers of $1/\phi$, we have,
\begin{equation}\label{bk-3}
u_0  = \phi _x ,v_0  =  - \phi _x \phi _y ,w_0  =  - \phi _x^2 ,v_1  = \phi _{xy} ,w_1  = \phi _{xx} .
\end{equation}

Substituting Eqs.(\ref{bk-2}-\ref{bk-3}) into system(\ref{bk-1}) and setting the coefficients of powers of $1/\phi^3$ to zero, can obtain two
\begin{equation}\label{bk-4}
\begin{array}{l}
 v_2  = u_{1y} , \\
 \phi _t  =  - 12\phi _x u_1^2  - 12\phi _x w_2  - 12\phi _{xx} u_1  - 12\phi _{xxx} , \\
 \end{array}
\end{equation}
where $u_1,v_2,w_2$  are seed solution of the (2+1) dimensional HBK system. From the above standard truncated Painlev\'{e} expansion system(\ref{bk-1}), we have the following BT theorem and nonlocal symmetry theorem.

\textbf{Theorem 1} If the function $\phi$ can be determined by the Eq.(\ref{bk-4}), then Eqs.(\ref{bk-2}) are just the solutions of the (2+1)-dimensional HBK system(\ref{bk-1}).

\textbf{Proof} By direct verification.

\textbf{Theorem 2}  The HBK system (1) has the residual symmetry given by
\begin{equation}\label{bk-5}
\sigma ^u  = \phi _x ,\sigma ^v  = \phi _{xy} ,\sigma ^w  = \phi _{xx} ,
\end{equation}
where $u, v,w$ and $\phi$ satisfy the non-auto BT.

\textbf{Proof} By direct verification.

To find out the symmetry group of the residual symmetry, study the Lie point symmetries of the whole prolonged equation system instead of the single system(\ref{bk-1}). From (\ref{bk-5}), it can be apparently seen that the nonlocal symmetry contains the space derivative of function $\phi$. Then, to localize the nonlocal symmetry (\ref{bk-5}), we introduce the following transformations.
\begin{equation}\label{bk-6}
\phi _x  = \phi _1 ,\phi _{xx}  = \phi _{1x}  = \phi _2 ,\phi _{xy}  = \phi _{1y}  = \phi _3 ,\phi _y  = \phi _4,
\end{equation}
and assume that the vector of the symmetries has the form,
\begin{center}
 $V = X\frac{\partial }{{\partial x}} + Y\frac{\partial }{{\partial y}} + T\frac{\partial }{{\partial t}} + U\frac{\partial }{{\partial u}} + V\frac{\partial }{{\partial v}} + W\frac{\partial }{{\partial w}} + \psi \frac{\partial }{{\partial \phi }} + \psi _1 \frac{\partial }{{\partial \phi _1 }} + \psi _2 \frac{\partial }{{\partial \phi _2 }} + \psi _3 \frac{\partial }{{\partial \phi _3 }},$
\end{center}
where $X, Y, T, U, V, W, \Psi, \Psi _1, \Psi _2, \Psi _3$ are the functions with respect to $x,y, t, u, v, w, \phi, \phi _1, \phi _2, \phi _3 $, which means that the closed system is invariant under the infinitesimal transformations.
\begin{center}
$(x,y,t,u,v,w,\phi ,\phi _1 ,\phi _2 ,\phi _3 ) \to (x + \varepsilon X,x + \varepsilon Y,x + \varepsilon T,x + \varepsilon U,...,x + \varepsilon \Psi _3 ),$
\end{center}
with
\begin{equation}\label{bk-7}
\begin{array}{l}
 \sigma ^u  = X\frac{\partial }{{\partial x}} + Y\frac{\partial }{{\partial y}} + T\frac{\partial }{{\partial t}} + U\frac{\partial }{{\partial u}},~~~~~~~~~~\sigma ^v  = X\frac{\partial }{{\partial x}} + Y\frac{\partial }{{\partial y}} + T\frac{\partial }{{\partial t}} + V\frac{\partial }{{\partial v}}, \\
 \sigma ^w  = X\frac{\partial }{{\partial x}} + Y\frac{\partial }{{\partial y}} + T\frac{\partial }{{\partial t}} + W\frac{\partial }{{\partial w}},~~~~~~~~\sigma ^\phi  = X\frac{\partial }{{\partial x}} + Y\frac{\partial }{{\partial y}} + T\frac{\partial }{{\partial t}} + \Psi \frac{\partial }{{\partial \phi }}, \\
 \sigma ^{\phi _1 }  = X\frac{\partial }{{\partial x}} + Y\frac{\partial }{{\partial y}} + T\frac{\partial }{{\partial t}} + \Psi _1 \frac{\partial }{{\partial \phi _1 }},~~~~~~\sigma ^{\phi _2 }  = X\frac{\partial }{{\partial x}} + Y\frac{\partial }{{\partial y}} + T\frac{\partial }{{\partial t}} + \Psi _2 \frac{\partial }{{\partial \phi _2 }}, \\
 \sigma ^{\phi _3 }  = X\frac{\partial }{{\partial x}} + Y\frac{\partial }{{\partial y}} + T\frac{\partial }{{\partial t}} + \Psi _3 \frac{\partial }{{\partial \phi _3 }},~~~~~~\sigma ^{\phi _4 }  = X\frac{\partial }{{\partial x}} + Y\frac{\partial }{{\partial y}} + T\frac{\partial }{{\partial t}} + \Psi _4 \frac{\partial }{{\partial \phi _4 }}, \\
 \end{array}
\end{equation}
moreover, $\sigma^u, \sigma^v, \sigma^w, \sigma^f, \sigma^{f_1}, \sigma^{f_2}, \sigma^{f_3}$ satisfy the linearized equations of (\ref{bk-1}), (\ref{bk-4}) and (\ref{bk-6}).

It is not difficult to verify that the solution of (\ref{bk-7}) has the form,
\begin{equation}\label{bk-7-1}
\begin{array}{l}
 X = F_6 (t) - xF_{2t} (t),Y = c_2  - F_5 (y),T = c_1  - 3F_2 (t), \\
 U = F_1 (y)\phi _1  + uF_{2t} (t),V = F_{1y} (y)\phi _1  + F_1 (y)\phi _3  + v(F_{2t} (t) + F_{5y} (y)), \\
 W = F_1 (y)\phi _2  - \frac{1}{{12}}(xF_{2tt} (t) - F_{6t} (t)) + 2wF_{2t} (t), \\
 \Psi  =  - F_1 (y)\phi ^2  + F_3 (y)\phi  + F_4 (y), \\
 \Psi _1  =  - 2F_1 (y)\phi \phi _1  + F_3 (y)\phi _1  + F_{2t} (t)\phi _1 , \\
 \Psi _2  =  - 2F_1 (y)\phi _1^2  + F_3 (y)\phi _2  - 2F_1 (y)\phi \phi _2  + 2F_{2t} (t)\phi _2 , \\
 \Psi _3  =  - 2F_1 (y)\phi _1 \phi _4  + F_3 (y)\phi _1  - 2F_1 (y)\phi _4 \phi _1  + (F_{2t} (t) + F_3 (y) + F_{5y} (y) - 2F_1 (y)\phi )\phi _3 , \\
 \Psi _4  =  - 2F_1 (y)\phi \phi _4  + F_3 (y)\phi  + F_{4y} (y) - F_1 (y)\phi ^2  + (F_3 (y) + F_{5y} (y))\phi _4,  \\
 \end{array}
\end{equation}

The results (\ref{bk-7-1}) show that the nonlocal symmetry (\ref{bk-5}) in the original space $(x,y,t,u,v,w)$ has been successfully localized to a Lie point symmetry in the enlarged space $(x,y,t,u,v,w,\psi,\psi_1 ,\psi_2 ,\psi_3 , \psi _4)$ .

After succeeding in making the nonlocal symmetry(\ref{bk-5}) equivalent to Lie point symmetry (\ref{bk-7-1}) of the related prolonged system, we can construct the explicit solutions naturally by Lie group theory. With the Lie point symmetry(\ref{bk-7-1}), For the sake of simplicity, let $F_1(y)=1,F_2 (t) = F_3 (y) = F_4 (y) = F_5 (y) = F_6 (t) = 0,$ by solving the following initial value problem,

\begin{equation}\label{bk-7-2}
\begin{array}{l}
 \frac{{d\bar u(\varepsilon )}}{{d\varepsilon }} = \bar \phi _1 (\varepsilon ),\frac{{d\bar v(\varepsilon )}}{{d\varepsilon }} = \bar \phi _3 (\varepsilon ),\frac{{d\bar w(\varepsilon )}}{{d\varepsilon }} = \bar \phi _2 (\varepsilon ),\frac{{d\bar \phi (\varepsilon )}}{{d\varepsilon }} =  - \bar \phi ^2 (\varepsilon ),\frac{{d\bar \phi _1 (\varepsilon )}}{{d\varepsilon }} =  - 2\bar \phi (\varepsilon )\bar \phi _1 (\varepsilon ), \frac{{d\bar \phi _2 (\varepsilon )}}{{d\varepsilon }} = \\
- 2\bar \phi _1^2 (\varepsilon ) - 2\bar \phi (\varepsilon )\bar \phi _2 (\varepsilon ),\frac{{d\bar \phi _3 (\varepsilon )}}{{d\varepsilon }} =  - 2\bar \phi _1 (\varepsilon )\bar \phi _4 (\varepsilon ) - 2\bar \phi (\varepsilon )\bar \phi _3 (\varepsilon ),\frac{{d\bar \phi _4 (\varepsilon )}}{{d\varepsilon }} =  - 2\bar \phi (\varepsilon )\bar \phi _4 (\varepsilon ) - \bar \phi ^2 (\varepsilon ), \\
 \bar u(\varepsilon )\left| {_{\varepsilon  = 0} } \right. = u,\bar v(\varepsilon )\left| {_{\varepsilon  = 0} } \right. = v,\bar w(\varepsilon )\left| {_{\varepsilon  = 0} } \right. = w,\bar \phi (\varepsilon )\left| {_{\varepsilon  = 0} } \right. = \phi ,\bar \phi _i (\varepsilon )\left| {_{\varepsilon  = 0} } \right. = \phi _i (i = 1,2,3,4), \\
 \end{array}
\end{equation}

Then the solution of the initial value problem (\ref{bk-7-2}) leads to the following group theorem for the enlarged system.

\textbf{Theorem 3}  If $\{u, v,w,\phi,\phi_1,\phi_2,\phi_3,\phi_4\}$ is a solution
of the prolonged system Eqs.(\ref{bk-1})and (\ref{bk-4}), so
\begin{center}
$\begin{array}{l}
 \bar u(\varepsilon ) = \frac{{\phi _1 }}{{\varepsilon \phi ^2  + \phi }} + \frac{{\phi _1  + \phi u}}{\phi },\bar v(\varepsilon ) = \frac{{2\phi _1 \phi _4  - \phi \phi _3 }}{{\varepsilon \phi ^3  + \phi ^2 }} - \frac{{\phi _1 \phi _4 }}{{\varepsilon ^2 \phi ^4  + 2\varepsilon \phi ^3  + \phi ^2 }} + \frac{{\phi ^2 v + \phi \phi _3  - \phi _1 \phi _4 }}{{\phi ^2 }}, \\
 \bar w(\varepsilon ) = \frac{{\phi ^2 w + \phi \phi _2  - \phi _1^2 }}{{\phi ^2 }} - \frac{{\phi _1^2 }}{{\varepsilon ^2 \phi ^4  + 2\varepsilon \phi ^3  + \phi ^2 }} + \frac{{2\phi _1^2  - \phi \phi _2 }}{{\varepsilon \phi ^3  + \phi ^2 }},\bar \phi (\varepsilon ) = \frac{\phi }{{1 + \varepsilon \phi }},\bar \phi _1 (\varepsilon ) = \frac{{\phi _1 }}{{(1 + \varepsilon \phi )^2 }}, \\
 \bar \phi _2 (\varepsilon ) = \frac{{\varepsilon \phi \phi _2  - 2\varepsilon \phi _1^2  + \phi _2 }}{{(1 + \varepsilon \phi )^3 }},\bar \phi _3 (\varepsilon ) = \frac{{\varepsilon \phi \phi _3  - 2\varepsilon \phi _1 \phi _4  + \phi _3 }}{{(1 + \varepsilon \phi )^3 }},\bar \phi _4 (\varepsilon ) = \frac{{\phi _4 }}{{(1 + \varepsilon \phi )^2 }}. \\
 \end{array}$
\end{center}
is also a solution of  Eqs.(\ref{bk-1})and (\ref{bk-4}).

Theorem 3 shows that the residual symmetry(\ref{bk-5}) coming from the truncated Painlev\'{e} expansion is just the infinitesimal form of the group(\ref{bk-7-1}).

\section{Exact solutions of 2+1-dimensional HBK System}

In order to give solutions of system (\ref{bk-1}), one should solve the Eqs.(\ref{bk-4}). But to find the general solution of (\ref{bk-4}) for any fixed $u_1,w_2$ is still quite difficult. Fortunately, one can verify that the seed solutions $u_1,w_2$ are arbitrary functions of $x$ and $t$.  Substituting $u_1  = u_1(x,t),w_2 = w_2(x,t)$ into Eqs.(\ref{bk-4}), we have
\begin{equation}\label{bk-8}
\begin{array}{l}
 v_2 = 0 , \\
 \phi _t  =  - 12\phi _x u_1^2  - 12\phi _x w_2  - 12\phi _{xx} u_1  - 12\phi _{xxx} , \\
 \end{array}
\end{equation}

In order to better understand the localized coherent structures of the (2+1) dimensional HBK system, we find it useful to apply the
variable separation method to this system. We assume that
\begin{equation}\label{bk-8-1}
\phi=a_1 p(x,t)+a_2 q(y,t),
\end{equation}
and $w_2$ is determining by the following form,
\begin{equation}\label{bk-9}
\begin{array}{l}
 w_2 =  - u_1^2  - \frac{1}{{12a_1 p_x }}(12a_1 p_{xx} u_1 + a_2 F_{1t}  + a_1 p_t  + 4a_1 p_{xxx} ), \\
 q = F_1+ F_2 , \\
 \end{array}
\end{equation}
where $F_1=F_1 (t),F_2=F_2 (y)$.

Substituting Eqs.(\ref{bk-3}),(\ref{bk-8-1}),(\ref{bk-9}) in to (\ref{bk-2})and get the exact solutions of the HBK system,

\begin{equation}\label{bk-10}
\begin{array}{l}
 u = \frac{{\phi _x }}{\phi } + u_1 , \\
 v =  - \frac{{\phi _x \phi _y }}{{\phi ^2 }} + \frac{{\phi _{xy} }}{\phi }, \\
 w =  - \frac{{\phi _x^2 }}{{\phi ^2 }} + \frac{{\phi _{xx} }}{\phi } + w_2 , \\
 \end{array}
\end{equation}
where $u_1$ are arbitrary functions of $x,t$, $\phi$ and $w_2$ are determined by (\ref{bk-8-1}),(\ref{bk-9}) respectively.

Because (\ref{bk-8-1}) contain the arbitrary function $p(x,t)$, so, there are abundant different structures to the solutions of (\ref{bk-10}). In this section, we focus on soliton solutions and periodic wave solutions .

\subsection {Soliton solution}

If we select $p(x,t)$ as some types of some smooth functions, we can construct exact soliton solutions of the 2+1 dimensional HBK system. For instance, if we select,
\begin{equation}\label{bk-11}
p = \sec h(\xi ),\xi  = x - \omega t,
\end{equation}
which leads to the single soliton solution of Eqs.(\ref{bk-1}). In order to study the properties of the solution, we plot the structure of the solution with $F_1(t)=1,F_2(y)=y,u_1=1,\omega=0.1$,

\subsection {Multiple resonant soliton solutions}
 For the next studies, we will seek the other type of soliton solutions, i.e. resonant soliton solution which has been widely studied. If we select
$p$ as,
\begin{equation}\label{bk-12}
p =  - \frac{1}{2}\ln \left[ {1 + \sum\limits_{i = 1}^n {\exp (k_i x + \omega _i t)} } \right],
\end{equation}
via the (\ref{bk-8-1}) along with the solution (\ref{bk-10}), the (n+1) resonant soliton solutions of Eqs. (\ref{bk-1}) can be directly obtained. In the nest section, we will seek various interaction solutions between different types of excitations.

\section{Interaction solutions for the HBK system}

For (2+1)-dimensional higher-order Broer-Kaup equations (\ref{bk-1}), the generalized tanh function expansion reads,
\begin{equation}\label{H-1}
\begin{array}{l}
 u = u_0  + u_1 \tanh (f), \\
 v = v_0  + v_1 \tanh (f) + v_2 \tanh ^2 (f), \\
 w = w_0  + w_1 \tanh (f) + w_2 \tanh ^2 (f), \\
 \end{array}
\end{equation}
where $f$ is an undetermined function of $x,y$ and $t$, and the expansion coefficient $u_0, u_1, v_0, v_1, v_2,w_0,w_1,w_2$ will be
determined by vanishing the coefficients of powers tanh($f$). Substituting expression (\ref{H-1}) into (\ref{bk-1}) yields,

\begin{equation}\label{H-2}
\begin{array}{l}
 u_1  = f_x , \\
 v_0  = f_x f_y  + u_{0y} ,v_1  = f_{xy} ,v_2  =  - f_x f_y , \\
 w_0  =  - 12f_x^{ - 1} (12u_0^2 f_x  - 8f_x^3  + 12u_0 f_{xx}  + f_t  + 4f_{xxx} ), \\
 w_1  = f_{xx} ,w_2  =  - f_x^2 , \\
 \end{array}
\end{equation}
and the function $f$ and $u_0$ only needs to satisfy,

\begin{equation}\label{H-3}
\begin{array}{l}
 u_{0xy}  =  - \frac{1}{{12}}f_x^{ - 2} (8f_{xy} f_x^3  + f_x f_{yt}  + 4f_x f_{xxxy}  - f_t f_{xy}  - 4f_{xxx} f_{xy}  + 24u_0 u_{0y} f_x^2  \\
  + 12u_{0y} f_x f_{xx}  + 12u_0 f_x f_{xxy}  - 12u_0 f_{xx} f_{xy} ), \\
 \end{array}
\end{equation}

\begin{equation}\label{H-4}
F(u_0,f)=0.
\end{equation}

Because the Eq.(\ref{H-4}) is very prolix which can be seen in appendix, here omitting it. It is quite difficult to find the general solution of $u_0,f$ of (\ref{H-3},\ref{H-4}). For the sake of simplicity, we select $u_0=0$, substituting the ansatz into (\ref{H-3}) we obtain,

\begin{equation}\label{H-5}
f_{xxxy}  = \frac{1}{4}f_x^{ - 1} (f_t f_{xy}  - f_x f_{yt}  - 8f_x^3 f_{xy}  + 4f_{xxx} f_{xy} ),
\end{equation}
substituting the (\ref{H-5}) into Eq.(\ref{H-4}), then Eq.(\ref{H-4}) simplified as,
\begin{equation}\label{H-6}
8f_{xxx} f_{xy}  + f_{xxx} f_{xx}  - 8f_{xy} f_x^3  + f_t f_{xy}  - f_x f_{yt}  = 0.
\end{equation}

The next work is to solve the Eq.(\ref{H-5}), we get the following three types of exact interaction solutions.

\textbf{Case 1: Variable separation solution}

It is not difficult to verify that Eq.(\ref{H-6}) possesses the following variable separation solution,
\begin{equation}\label{H-7}
f = f_1 (x) + f_2 (y) + f_3 (t),
\end{equation}
which leads to the interaction solution of Eqs.(\ref{bk-1}),
\begin{center}
$\begin{array}{l}
 u = f_{1x} \tanh (f), \\
 v = f_{1x} f_{2y}  - f_{1x} f_{2y} \tanh ^2 (f), \\
 w = \frac{1}{{12}}f_{1x}^{ - 1} (8f_{1x}^3  - 4f_{1xxx}  - f_{3t}  + 12f_{1x} f_{1xx} \tanh (f) - 12f_{1x}^3 \tanh ^2 (f)). \\
 \end{array}$
\end{center}

\textbf{Remark 1:} Due to the arbitrariness of function $f_1(x), f_2(y), f_3(t)$, we are able to construct many types of exact interaction solutions. For instance, if we select $f_1(x)=k_1 x, f_2(y)=k_2 y, f_3(t)=k_3 t$, then the exact single soliton solution of the 2+1 dimensional HBK system can be obtained. If choose other type functions to $f_1(x), f_2(y), f_3(t)$, we will get more interaction solutions of HBK system.

\textbf{Case 2: The first type of special soliton-cnoidal waves solution}

By solving Eq.(\ref{H-6}), we just obtain a special solution in the form,
\begin{equation}\label{H-8}
f = sn(k_1 x + k_2 t,m_1 ),
\end{equation}
where $k_1,k_2$ are arbitrary constants and $m_1$ is modulus of Jacobi elliptic function. Then the exact interaction solution of the 2+1 dimensional HBK system can be obtained in the form,
\begin{equation}\label{H-8-1}
\begin{array}{l}
 u = k_1 CD\tanh (S),~~~~v = 0, \\
 w =  - k_1^2 C^2 D^2 \tanh ^2 (S) - k_1^2 S(m^2 C^2  + D^2 )\tanh (S) - \frac{1}{{12k_1 }}\\
 (k_2  + 16k_1^3 m_1^2 S^2  - 8k_1^3 C^2 D^2  - 4k_1^3 m_1^2 C^2  - 4k_1^3 D^2 ), \\
 \end{array}
\end{equation}
where $S \equiv sn(k_1 x + k_2 t,m_1 ),C \equiv cn(k_1 x + k_2 t,m_1 ),D \equiv dn(k_1 x + k_2 t,m_1 )$.

The solution given in (\ref{H-8-1}) denotes the analytic interaction solution between the soliton and the cnoidal periodic wave.

\textbf{Case 3: The second type of special soliton-cnoidal waves solution}

We just write a special solution of the equation Eq.(\ref{H-6}) in the form,
\begin{equation}\label{H-8-2}
f = l_0 x + l_1 y + l_2 t + c{\rm{F}}(sn(\omega _0 x + \omega _1 y + \omega _2 t,m_2),m_2),
\end{equation}
With the help of the Maple, substituting expression (\ref{H-8-2}) into Eq.(\ref{H-6}) yields,
\begin{equation}\label{H-9}
f = l_0 x + l_1 y + \frac{1}{{\omega _0 }}(8c^3 \omega _0^4  + 24c^2 \omega _0^3 l_0  + 24cl_0^2 \omega _0^2  + 8l_0^3 \omega _0  + l_0 \omega _2 )t + c{\rm{F}}(sn(\Delta _3,m_2),m_2),
\end{equation}
$l_0,l_1,\omega _0,\omega _1,\omega _2$ and $c$ are arbitrary constants and $m_2$ is modulus of Jacobi elliptic function. Substituting (\ref{H-9}) into (\ref{H-1}) and the exact interaction solution of the HBK system can be obtained in the form,

\begin{equation}\label{H-10}
\begin{array}{l}
 u_0  = v_1  = w_1  = 0,\\
 u_1  = ({{c\omega _0 DC + l_0 \Delta _1 \Delta _2 )} \mathord{\left/
 {\vphantom {{c\omega _0 DC + l_0 \Delta _1 \Delta _2 )} {(\Delta _1 \Delta _2 )}}} \right.
 \kern-\nulldelimiterspace} {(\Delta _1 \Delta _2 )}}, \\
 v_0  = (l_0 l_1  + l_0 l_1 m_2^2 S^4  - l_0 l_1 m_2^2 S^2  - l_0 l_1 S^2  + c\omega _0 l_1 DC\Delta _1 \Delta _2  + c\omega _1 l_0 DC\Delta _1 \Delta _2  +  \\
 c^2 \omega _0 \omega _1 m_2^2 S^4  - c^2 \omega _0 \omega _1 m_2^2 S^2  - c^2 \omega _0 \omega _1 S^4  + c^2 \omega _0 \omega _1 )/(m_2^2 S^4  - m_2^2 S^2  - S^2  + 1), \\
 w_2  =  - (l_0^2  + c^2 \omega _0^2  + c^2 \omega _0^2 m_2^2 S^4  - c^2 \omega _0^2 m_2^2 S^2  - c^2 \omega _0^2 S^2  + 2c\omega _0 l_0 DC\Delta _1 \Delta _2  +  \\
 l_0^2 m_2^2 S^4  - l_0^2 m_2^2 S^2  - l_0^2 S^2 )/(m_2^2 S^4  - m_2^2 S^2  - S^4  + 1), \\
 v_2  =  - v_0 , \\
 w_0  = \frac{1}{{12}}(8c^3 \omega _0^4 CD - 24cl_0^2 \omega _0^2 \Delta _1 \Delta _2  - c\omega _0 \omega _2 CD + 24cl_0^2 \omega _0^2 CD - l_0 \omega _2 \Delta _1 \Delta _2  \\
  - 8c^3 \omega _0^4 \Delta _1 \Delta _2 )/(c\omega _0^2 CD + l_0 \omega _0 \Delta _1 \Delta _2 ), \\
 \end{array}
\end{equation}
where $S \equiv sn(\Delta _3 ,m),C \equiv cn(\Delta _3 ,m),D \equiv (\Delta _3 ,m),\Delta _1  = \sqrt {1 - sn^2 (\Delta _3 ,m)} ,\Delta _2  = \sqrt {1 - sn^2 (\Delta _3 ,m)m^2 } ,\Delta _3  = \omega _0 x + \omega _1 y + \omega _2 t$.

In solution (\ref{H-10}) ,  $F(\xi,m)$ is the first type of incomplete elliptic integral, i.e. $F(\xi ;m) = \int_0^\xi  {\frac{{dt}}{{\sqrt {(1 - t^2 )(1 - m^2 t^2 )} }}}.$

\textbf{Case 4: The third type of special soliton-cnoidal waves solution}

If we assume that Eq.(\ref{H-6}) equation has the following form solution,
\begin{equation}\label{H-11}
f = \lambda _0 x + \lambda _1 y + \lambda _2 t + \mu{\rm{E}}(sn(\gamma _0 x + \gamma _1 y + \gamma _2 t,m_3),m_3),
\end{equation}
where $E(\xi,m)$ is incomplete elliptic integrals of the second kind,i.e. ${\rm{E}}(\xi ,m) = \int_0^\xi  {\frac{{\sqrt {1 - m^2 t^2 } }}{{\sqrt {1 - t^2 } }}dt}$. It is not difficult to solve above coefficients by substituting (\ref{H-11}) into Eq.(\ref{H-6}) which have following two nontrivial solutions,
\begin{equation}\label{H-12}
\begin{array}{l}
 \{ \lambda _0  = \lambda _0 ,\lambda _1  = \lambda _1 ,\lambda _2  = \lambda _2 ,\mu  = \mu ,\gamma _0  = 0,\gamma _1  = \gamma _1,\gamma _2  = \gamma _2 ,m_3  = m_3 \} , \\
 \{ \lambda _0  = \lambda _0 ,\lambda _1  = \lambda _1 ,\lambda _2  = \lambda _2 ,\mu  = \mu ,\gamma _0  = \gamma _0,\gamma _1  = 0 ,\gamma _2  = \gamma _2 ,m_3  = m_3  \} , \\
 \end{array}
\end{equation}
by substituting (\ref{H-11}) and (\ref{H-12}) into (\ref{H-1}), the exact interaction solution of the HBK system can be obtained. Because the results are similar to case 3, so here omitting it.

\textbf{Remark 2:} Interaction solution between the solitary wave and the cnoidal wave when the value of the Jacobi elliptic function modulus is not equal to 1.  This kind of solution can be easily applicable to the analysis of physically interesting processes.

\textbf{Remark 3:} Using the theorem 3 and the results of above two sections, we can construct more group invariant solutions of (2+1) dimensional HBK system.

\section{Discussion and Summary}

In summary, using the nonlocal symmetry method, the residual symmetries of (2+1)-dimensional higher order Broer-Kaup system can be localized to Lie point symmetries after introducing suitable prolonged systems, and symmetry groups can also be obtained from the Lie point symmetry approach via the localization of the residual symmetries. By developing the truncated Painlev\'{e} analysis, we using the CTE method to solve the HBK system. It is found that the HBK system is not only integrable under some nonstandard meaning but also CTE solvable. Some interaction solutions among solitons and other types of nonlinear waves which may be explicitly expressed by the Jacobi elliptic functions and the corresponding elliptic integral are constructed. To leave it clear, we give out four types of soliton+cnoidal periodic wave solution. More types of these soliton-cnoidal wave solutions need our further study.

\section*{Acknowledgments}

This work is supported by National Natural Science Foundation of China under Grant (Nos.11505090,11171041,\\
11405103, 11447220), Research Award Foundation for Outstanding Young Scientists of Shandong Province \\(No.BS2015SF009).

\newpage

\textbf{Appendix}

$ F(u_0,f)= 12u_{0yt} f_x^4 - 28u_0 f_x f_{xy} f_t f_{xx}  - 544u_0 f_x f_{xx} f_{xxx} f_{xy}   - 64f_x^7 f_{xy}  - 8f_x^5 f_{yt}  - 4f_x^3 f_{xxyt}  + 80f_x^5 f_{xxxy}+ 16f_x^4 f_{xy} f_t  + 96u_{0y} f_x^5 f_{xx}  + f_t f_x^2 f_{ty}  - f_x^2 f_{xy} f_x  + 64u_0^2 f_x^5 f_{xy}  + 8u_0^2 f_x^3 f_{ty}  + 96u_0^3 f_x^3 f_{xxy}  - 16u_0^2 f_x^3 f_{xxxy}+ 160u_0 f_x^5 f_{xxy}  - 4u_0 f_x^3 f_{xyt}  + 320u_{0x} f_x^5 f_{xy}  + 4u_{0x} f_x^3 f_{yt}  - 12u_{0y} f_x^3 f_{xt}  - 64f_x^3 f_{xy} f_{xx}^2  - 96u_{0y} f_x^4 u_{0xx}- 48u_{0xx} f_x^3 f_{xxy}  - 80u_{0x} f_x^3 f_{xxxy}  - 64u_0 f_x^3 f_{xxxxy}  - 96u_{0y} f_x^3 f_{xxxx}  + 256f_x^4 f_{xx} f_{xxy}  + 240f_x^4 f_{xxx} f_{xy}  - 16f_x^2 f_{xxx} f_{yt}+ 12f_x^2 f_{xx} f_{xyt}  + 48f_x^2 f_{xxxxy} f_{xx}  + 8f_x^2 f_{xxxy} f_t  + 8f_x^2 f_{xxy} f_{xt}  + 4f_x^2 f_{xy} f_{xxt}  + 16f_x^2 f_{xxxxx} f_{xy}  + 32f_x^2 f_{xxxx} f_{xxy}- 24f_x f_{xx}^2 f_{yt}  - 96f_x f_{xx}^2 f_{xxxy}  - 288f_x f_{xx}^3 u_{0y}  + 44f_t f_{xx}^2 f_{xy}  + 176f_{xxx} f_{xx}^2 f_{yx}  + 528u_0 f_{xx}^3 f_{yx}  + 80f_{xxx} f_x^2 f_{yxxx}+ 288u_0^2 f_x^3 f_{xx} u_{0y}  + 16u_0^2 f_x^2 f_{xy} f_{xxx}  + 192u_0^2 f_x^2 f_{xx} f_{xxy}  - 4u_{0x} f_x^2 f_{xy} f_t  + 416u_0 f_x^4 f_{xy} f_{xx}  + 96u_0 u_{0y} f_x^3 f_{xxx}+ 240u_0 f_x^2 f_{xx} f_{xxxy}  + 48u_{0xx} f_x^2 f_{xx} f_{xy}  + 80u_{0x} f_x^2 f_{xy} f_{xxx}  + 64u_0 f_x^2 f_{xy} f_{xxxx}  + 304u_0 f_x^2 f_{xxx} f_{xxy}  + 384u_{0y} f_x^2 f_{xx} f_{xxx}- 20f_x f_{xx} f_{xxy} f_t  - 20f_x f_{xx} f_{xy} f_{xt}  - 80f_x f_{xx} f_{xxxx} f_{xy}  - 80f_x f_{xx} f_{xxx} f_{xxy}  + 240u_{0x} f_{xx} f_{xxy} f_x^2  - 240u_{0x} f_x f_{xy} f_{xx}^2- 528u_0 f_x f_{xxy}f_{xx}^2  - 24f_x f_{xy} f_{xxx} f_t  + 24u_{0y} f_t f_{xx} f_x^2  + 12u_0 f_{yt} f_{xx} f_x^2  - 192u_0^2 f_x f_{xy} f_{xx}^2  + 16u_0 f_t f_{xxy} f_x^2 + 4u_0 f_{xt} f_{xy} f_x^2  - 8u_0^2 f_t f_{xy} f_x^2  - 192u_0 u_{0y} f_x^6  + 192u_0^3 u_{0y} f_x^4  + 24u_0 u_{0y} f_x^3 f_t -16 f_x^{3} f_{xxxxxy}=0, $

%
%
%
%

\small{
}
\end{document}